\documentclass[
aps,
prd,
12pt,
nopreprintnumbers,
showpacs,
eqsecnum,
nofootinbib
]{revtex4-1}

\usepackage{graphicx}
\usepackage{amssymb}

\begin{document}

\title{An ostentatious model of cosmological scalar-tensor theory}
\author{Nahomi Kan}\email[]{kan@gifu-nct.ac.jp}
\affiliation{National Institute of Technology, Gifu College,
Motosu-shi, Gifu 501-0495, Japan
}
\author{Kiyoshi Shiraishi}\email[]{shiraish@yamaguchi-u.ac.jp}
\affiliation{
Graduate School of Sciences and Technology for Innovation, Yamaguchi
University, Yamaguchi-shi, Yamaguchi 753--8512, Japan}
\date{\today}

\begin{abstract}
We consider a novel model of gravity with a scalar field described by the
Lagrangian with higher order derivative terms in a cosmological context.  The
model has the same solution for the homogeneous and isotropic
universe as in the model with the Einstein gravity, notwithstanding the additional
higher order terms.  A possible modification scenario is briefly discussed lastly.
\end{abstract}



\maketitle

\section{Introduction}
\label{sec1}

Basics of modern cosmology was born almost at the time when Einstein discovered his
theory of gravitation, called the general theory of relativity, about one hundred
years ago. The most outstanding fact is that general relativity has passed
repetitive tests with high degrees of accuracy until very recently.
Nevertheless, modification of general relativity or modified theory of gravity is
eagerly discussed by many authors
\cite{NO1,SF,FT,NO2,CL,CFPS,Koyama,NOO,Heisenberg}, who are willing naturally to
explain the acceleration of our universe in the present epoch
\cite{darkenergy1,darkenergy2} and in the very early universe
\cite{inflation}. To achieve the accelerating behavior, scalar or other degrees of
freedom are incorporated into the sector of Lagrangian that describes gravitation.

Another motivation of exploring alternative theories of gravity is found in the
theoretical pursuit of a  possible consistent quantum field theory of gravity.
Because the Planck mass has physical dimension, general relativity cannot
be renormalisable straightforwardly as a quantum field theory.
At least, inclusion of higher order terms of curvature tensors in the Lagrangian
is needed to control the UV behavior of the theory \cite{Stelle1}.

The recent observation of gravitational waves from a neutron star merger
restricts the difference in the velocity of the gravitational wave
and the light speed \cite{BBFLNS,CV,SJ,EZ} (see also earlier discussions
\cite{LT,LL,MLP}). Since this experimental limit is very severe, we accept giving
up many modified models of gravity, or very unnatural fine tuning in the theories.

More recently, Motohashi and Minamitsuji \cite{MM} proposed a possible form of the
Lagrangian for modified gravity, which leads to the exact general relativistic
solutions in a certain limit. In such models, the gravitational wave propagates
with the light speed.

Now, let us consider a cosmological models with a neutral scalar
field.
Suppose the following action of Einstein--scalar system:
\begin{equation}
I_0=\int d^4x\sqrt{-g}L_0=\int
d^4x\,\sqrt{-g}\,\left[R-\sigma\frac{1}{2}(\nabla\phi)^2-
V(\phi)\right]
\,,
\end{equation}
where $R$ is the Ricci
scalar, $(\nabla\phi)^2=g^{\mu\nu}\nabla_\mu\phi\nabla_\nu\phi$, and
$V(\phi)$ is the potential of the scalar field $\phi$.  If the constant $\sigma$
takes one, the kinetic term is a canonical one and on the other hand, if
$\sigma=-1$, the scalar is called phantom
\cite{phantom}. By taking the variation with respect to the metric, we find the
Einstein equation
\begin{equation}
T_{\mu\nu}=0\,
\end{equation}
with
\begin{eqnarray}
T_{\mu\nu}&\equiv&\frac{1}{\sqrt{-g}}\frac{\delta I_0}{\delta
g^{\mu\nu}}=R_{\mu\nu}-\frac{1}{2}Rg_{\mu\nu}
-\sigma\frac{1}{2}\nabla_{\mu}\phi\nabla_\nu\phi+\sigma\frac{1}{4}(\nabla\phi)^2
g_{\mu\nu}+\frac{1}{2}V(\phi)g_{\mu\nu}\nonumber \\
&\equiv&\tau_{\mu\nu}-\frac{1}{2}\tau g_{\mu\nu}+\frac{1}{2}V(\phi)g_{\mu\nu}\,,
\end{eqnarray}
where $R_{\mu\nu}$ is the Ricci tensor, $\tau_{\mu\nu}\equiv
R_{\mu\nu}-\sigma\frac{1}{2}\nabla_\mu\phi\nabla_\nu\phi$, and
$\tau\equiv\tau_\mu^\mu$.  Now, we consider a new Lagrangian
\begin{equation}
L=L_0+F(T_{\mu\nu})\,,
\end{equation}
yields the equation of motion which is satisfied with $T_{\mu\nu}=0$,
provided that the function $F$ is at least the second order or higher in
$T_{\mu\nu}$. The existence of the general relativistic solution means that
the existence of a massless graviton moving with the speed of light. 

The Lagrangian of this simple model  has apparently higher order terms in
curvature tensors, i.e., the model describes a higher-derivative theory.
A generic higher-derivative gravity contains ghost-like massive modes
as a perturbative field theory
\cite{Stelle1,Stelle2,BC,HJS,Alvarez,Schmidt,AKKLR,AP,Salvio}. Although this
fact may be a fault of the model, discussions have been made to unravel a mystery
in astrophysics, such as the problem with galactic rotation curves, by considering
Yukawa-like potential of the form $e^{-mr}/r$ associated with the massive modes
\cite{SS}. In addition, much theoretical interest has been found in study of
higher order gravities in diverse branches 
\cite{GT,Nelson1,Nelson2,SGT,LP,DLLP,LPP,LPPS1,LPPS2,HS,HZ,LPPS3,KKZ,LLW,LLL,GM,BCR}.

Modern theoretical cosmologists also hate a related ghost, so-called
Ostrogradsky ghost \cite{Ostrogradsky,Woodard}, which appears in the canonical
equation of motion of the scale factor and homogeneous fields in cosmology. Even if
we restrict ourselves to considering isotropic and homogeneous cosmological models
(or minisuperspace), the Lagrangian that contains the higher order terms in the the
second time derivatives yields the multiple degrees of freedom, which behave as
ghosts. In the simplest way to avoid the Ostrogradsky ghost in the metric theory
of gravity, we have to choose the combination of curvature tensors in the
Lagrangian to contain only linear order of the second time derivative of the scale
factor of the universe. Such an elaborated choice has been already known for the 
FLRW cosmology
\cite{MO1,MO2} and has been studied by the present authors
\cite{KKS1,KKS2,KS}. Thus, in this Letter, we study the model of higher order
scalar-tensor theory, which has the general relativistic solution in minisuperspace
governed by the field equation which includes at most the second time derivative
of the scale factor and the scalar field.

We are now standing at an intermediate position between modern
theoretical cosmologists and high energy theorists. We think that the
ghost-like fluctuations around the classical homogeneous background fields would
yield some physical implication if we can treat them correctly. The modification
of the theory will be discussed later in this Letter.

\section{action with higher order curvature terms}
\label{ost}

We shall begin with introducing the Meissner--Olechowski gravity in four
dimensional spacetime. The $n$-th order Meissner--Olechowski density is defined
here by using the Schouten tensor and the generalized Kronecker delta as
\cite{MO1,MO2,KKS1,KKS2,KS}
\begin{equation}
L_{MO}^{(n)}=-\delta^{\mu_1\cdots\mu_n}_{\nu_1\cdots\nu_n}
{S_R}^{\nu_1}_{\mu_1}\cdots{S_R}^{\nu_n}_{\mu_n}\equiv-[S_R\cdots S_R]
\,,
\end{equation}
where the symbol ${S_R}_\mu^\nu$ denotes the Schouten tensor (in our definition,
which is different from the original definition by a factor $1/2$ in four
dimensions), which is 
$S_R^{\mu\nu}=R^{\mu\nu}-\frac{1}{6}Rg^{\mu\nu}$ in four dimensions,
and the generalized Kronecker delta
$\delta^{\mu_1\mu_2\cdots\mu_p}_{\nu_1\nu_2\cdots\nu_p}$ is defined as
\begin{equation}
\delta^{\mu_1\mu_2\cdots\mu_p}_{\nu_1\nu_2\cdots\nu_p}
\equiv\left|\begin{array}{cccc}
\delta^{\mu_1}_{\nu_1} & \delta^{\mu_1}_{\nu_2} & \cdots &
\delta^{\mu_1}_{\nu_p} \\
\delta^{\mu_2}_{\nu_1} & \delta^{\mu_2}_{\nu_2} & \cdots &
\delta^{\mu_2}_{\nu_p} \\
\vdots & \vdots & \ddots & \vdots \\
\delta^{\mu_p}_{\nu_1} & \delta^{\mu_p}_{\nu_2} & \cdots &
\delta^{\mu_p}_{\nu_p} 
\end{array}\right|\quad (p=2,3,4)\,.
\end{equation}

The Lagrangian
density that consists of a linear combination of the Meissner--Olechowski density
includes at most linear order terms in the second derivative of the metric tensor
in the FLRW universe \cite{MO1,MO2,KKS1,KKS2,KS}.
Note that $L_{MO}^{(2)}=R_{\mu\nu}R^{\mu\nu}-\frac{1}{3}R^2$ in four dimensions.
The theory of Meissner--Olechowski gravity perturbatively describes a ghost-like
massive spin-2 mode in addition to a massless spin-2 graviton and contains no
scalar mode.

For the present purpose, we can consider the 
terms $[SS]$, $[SSS]$ and $[SSSS]$ in addition to $L_0$ in the Lagrangian, where
\begin{equation}
S_{\mu\nu}\equiv T_{\mu\nu}-\frac{1}{3}Tg_{\mu\nu}
\,,
\end{equation}
in four dimensions.
These  candidate terms, however, yields $O({h_{\mu\nu}}^2)$, $O({h_{\mu\nu}}^3)$
and $O({h_{\mu\nu}}^4)$ contributions in terms of the metric fluctuation
$h_{\mu\nu}=g_{\mu\nu}-\bar{g}_{\mu\nu}$, if the background metric
$\bar{g}_{\mu\nu}$ satisfies the Einstein equation
$T_{\mu\nu}=0$. Therefore, we consider here the Lagrangian
\begin{equation}
L=\alpha L_0+\beta [SS]+\gamma [S_\tau SS]+\delta [S_\tau S_\tau SS]\,,
\quad(\alpha, \beta, \gamma \mbox{ and } \delta \mbox{ are constants})
\label{o}
\end{equation}
with ${S_\tau}_{\mu\nu}\equiv\tau_{\mu\nu}-\frac{1}{6}\tau g_{\mu\nu}$,
which has the $O({h_{\mu\nu}}^2)$ contribution in each term and
reduces to the Meissner--Olechowski gravity in the limit of vanishing $\phi$ and
$V(\phi)$.
Because the explicit form of the full Lagrangian in terms of the individual field
content is lengthy, we only exhibit $\alpha L_0+\beta [SS]$ below:
\begin{eqnarray}
L&=&\alpha\left[R-\sigma\frac{1}{2}(\nabla\phi)^2-
V(\phi)\right]
-\beta\left[\left(R_\mu^\nu-\sigma\frac{1}{2}\nabla_\mu\phi\nabla^\nu\phi\right)^2-
\frac{1}{3}\left(R-\sigma\frac{1}{2}(\nabla\phi)^2\right)^2\right.\nonumber
\\ &
&\left.\qquad+\frac{1}{3}V(\phi)\left(R-\sigma\frac{1}{2}(\nabla\phi)^2
\right)-\frac{1}{3}V^2(\phi)\right]\,.
\label{ours}
\end{eqnarray}

We find that the expansion of the Lagrangian (\ref{o}) around the background
metric $\bar{g}_{\mu\nu}$ satisfying the field equation $T_{\mu\nu}=0$ yields
\begin{equation}
\sqrt{-g}L=\sqrt{-\bar{g}}\left[\alpha
\frac{1}{4}h^{\mu\nu}\bar{\nabla}^2h_{\mu\nu}-\left\{\beta+\gamma\frac{V(\bar{\phi})
}{3}+\delta\frac{V^2(\bar{\phi})}{18}\right\}\frac{1}{4}h^{\mu\nu}
\bar{\nabla}^2\bar{\nabla}^2h_{\mu\nu}+\cdots\right]\,,
\end{equation}
where $\bar{\phi}$ is the classical homogeneous solution of the general
relativistic field equation and $\bar{\nabla}^2$ denotes the covariant
d'Alembertian in terms of the background metric $\bar{g}_{\mu\nu}$. Thus, the
present model theory described the Lagrangian (\ref{o}) has the ghost-like massive
spin-2 mode of mass squared
$m^2(\bar{\phi})=\alpha(\beta+\gamma{V(\bar{\phi})
}/{3}+\delta{V^2(\bar{\phi})}/{18})^{-1}$, if the time evolution of the scalar
field is sufficiently slow. This mode causes Yukawa-type static potential
$\sim e^{-m(\bar{\phi})r}/r$ \cite{Stelle2}.
This field-dependent potential is interesting from an astrophysical perspective,
which can be concerned with the galactic length scale, as an extension of the
discussion in \cite{SS}.

So far, we have constructed an ostentatious model of higher order scalar-tensor
gravity, which has the general relativistic solution.
This model has a massless graviton as well as a ghost-like massive spin-2 mode,
which may affect some astrophysical phenomena.

In the first place, however, can we find the modification or new perspective in
cosmological evolution? How can we tell our model from general
relativity in cosmology at a large scale? We will serve an answer later, and we
first illustrate a possible moderate modification of our present model.

\section{a not-so-ostentatious cosmological model of gravity}
\label{nso}

We start with the simplest Lagrangian $L=\alpha L_0+\beta [SS]$, i.e.,
the case $\gamma=\delta=0$. One can seen that $L$ includes the pure higher
order term of curvature tensors $R_\nu^\mu R_\mu^\nu-\frac{1}{3}R^2$. 
As far as we consider the FLRW cosmological model, this term
 has no effect on the equation for the scale factor,
because the FLRW metric is conformally flat \cite{Ibison} and the term is
just the Weyl tensor squared modulo the Euler density in four dimensions.
Thus, we consider a modified model described by the Lagrangian density
$L'=L+\beta \left(R_\nu^\mu R_\mu^\nu-\frac{1}{3}R^2\right)$, which still leads to
the same FLRW solution as in the Einstein gravity described by $I_0$, aside from
the frame dependence. More explicitly, the modified Lagrangian can be written as
\begin{eqnarray}
L'&=&\alpha\left[R-\sigma\frac{1}{2}(\nabla\phi)^2-
V(\phi)\right]+\beta'\left[-\sigma\frac{1}{6}V(\phi)(\nabla\phi)^2
+\sigma^2\frac{1}{6}[(\nabla\phi)^2]^2
-\frac{1}{3}V^2(\phi)\right.\nonumber
\\ &
&\left.\quad+\frac{1}{3}\left(V(\phi)+\sigma(\nabla\phi)^2
\right)R-\sigma R^{\mu\nu}\nabla_\mu\phi\nabla_\nu\phi\right]
\,,
\label{ours2}
\end{eqnarray}
where we use $\beta'\equiv-\beta$ for convenience.  Note
that if
$\beta'\ge 0$, the total scalar potential $V+\beta'V^2/3$ can be positive
definite. Incidentally, this model can be regarded
as an extension of the $f(R,T,R_{\mu\nu}T^{\mu\nu})$ gravity \cite{OS,HHLSS}. The
model Lagrangian (\ref{ours2}) does not yield the spin-2 ghost, because the higher
derivative terms for the metric has been discarded. The massless mode, however,
does not propagate with the speed of light in this model described by
$L'$. To see this, we follow the method used in Refs.~\cite{KYY,FT1,FT2}. Thus, we
assume the line element with the perturbation $\Phi$:
\begin{equation}
ds^2=-dt^2+a^2(t)(1+2\Phi)d{\mathbf x}^2\,.
\end{equation}
Then, the action can be read as
\begin{equation}
\int d^4x\sqrt{-g}L'=\int d^4x a^3\left[\frac{w_4}{a^2}(\partial_i\Phi)^2
-3 w_1\dot{\Phi}^2+\cdots\right]\,,
\end{equation}
with
\begin{equation}
w_4=2\left(\alpha+\frac{\beta'}{3}V(\bar{\phi})\right)-\frac{2}{3}
\sigma\beta'\dot{\bar{\phi}}^2\,,\quad
w_1=2\left(\alpha+\frac{\beta'}{3}V(\bar{\phi})\right)+\frac{4}{3}
\sigma\beta'\dot{\bar{\phi}}^2\,,
\end{equation}
where the dot denotes the time derivative.
The speed of propagation of gravitational wave $c_{GW}$ is then given by
\cite{KYY,FT1,FT2}
\begin{equation}
c_{GW}^2=\frac{w_4}{w_1}=1-\frac{2\sigma\beta'}{w_1}\dot{\bar{\phi}}^2\,.
\end{equation}
The stringent bound from the observation of GW170817 $|c_{GW}^2-1|<10^{-15}$
\cite{BBFLNS} implies very strict fine tuning in this model parameters.

Note that further removing the non-minimal terms made of the product of the
curvature tensors and the scalar field and its derivatives makes the model very
similar to the energy-momentum-squared gravity \cite{RS,AKK,BB,ABCEK}.
Unfortunately, the renovated model obtained by the further removal of the
terms cannot have the same general relativistic solution as previously.

Note also that the present model can be considered as a model of induced gravity
\cite{Zee,Smolin,Adler}, if we consider the limit $\alpha\rightarrow 0$;
nevertheless, a classical FLRW solution is given by the solution of the equation
of motion derived from
$I_0$. In an academic and theoretical perspective, this is an interesting
possibility and is worth studying further.

Next, we will turn to consider the modified scenario of the higher order model
including higher curvature terms, after summarizing discussions.

\section{Summary and discussion -- a possible ``modification'' of the model}
\label{dis}

We have proposed a seemingly complicate and ostentatious model of higher order
scalar-tensor theory with general covariance, whose solution is given by that of
the simple Einstein gravity. The model generally has massive spin-2 modes in
addition to a massless graviton mode. Although, in the simplest case, removing the
massive mode is possible by discarding pure curvature polynomials, this omission
bring about the change in the speed of the massless graviton mode.

The conformally non-flat metric, especially the metric around compact objects is
the object concerned with future study in the present higher order model. The
extension of our model can be studied by incorporating the ideas of using
$F(T_{\mu\nu},\phi,\nabla^n\phi,\ldots)$, induced gravity, quantum cosmology,
higher-dimensional extension, and supersymmetry, etc.

Now, we will go back to the first place, and ask ourselves:
where is the ``modified'' gravity in our higher derivative model,
in which the same solution holds as in the Einstein gravity?
We propose a scenario of ``ghost condensation'' here.
Let us consider the simplest model with a massive spin-2 mode, which is governed
by the Lagrangian
\begin{equation}
L=\alpha L_0-\beta\left(T^{\mu\nu}T_{\mu\nu}-\frac{1}{3}T^{2}\right)\,.
\end{equation}
This Lagrangian can be replaced by an equivalent Lagrangian, using an 
auxiliary field $\tilde{S}_{\mu\nu}$ as \cite{KKS2,KS,HOW1,HOW2,HinS}
\begin{equation}
\tilde{L}=\alpha
L_0-\beta(2T_{\mu\nu}\tilde{S}^{\mu\nu}-\tilde{S}_{\mu\nu}\tilde{S}^{\mu\nu}
+\tilde{S}^2)\,,
\end{equation}
where $\tilde{S}\equiv\tilde{S}^\mu_\mu$.
Note that the Fierz--Pauli equation \cite{FP} for $\tilde{S}_\mu^\nu$ with mass
squared $\alpha/\beta$ can be derived from $\tilde{L}$ for a small fluctuation
around $\tilde{S}_\mu^\nu=0$ \cite{KS}.
 
We conjecture that the ghost-like tensor mode condensate, most simply as
$\langle\tilde{S}_\mu^\nu\rangle=\Lambda\delta_\mu^\nu$.
Then, the Lagrangian $\tilde{L}$ becomes
\begin{equation}
\tilde{L}=(\alpha+2\beta\Lambda)\left(R-\frac{1}{2}\sigma(\nabla\phi)^2
\right)-(\alpha+4\beta\Lambda)V(\phi)-\beta\Lambda^2\,,
\end{equation}
which apparently contains the cosmological term.
Therefore, the condensation of $\tilde{S}_\mu^\nu$ bring about the modification of
the whole theory.
We should add a comment, however.
For a positive cosmological constant, $\beta$ should be positive.
To obtain a ``natural'' condensation, the mass squared $\alpha/\beta$
around $\tilde{S}_\mu^\nu=0$ might be negative for obtaining
$\langle\tilde{S}_\mu^\nu\rangle\ne 0$. Thus, detailed investigation for the
mechanism of the ghost condensation are needed, with incorporating higher order
terms and/or quantum dynamics of the scalar and tensor fields in the flat or the de
Sitter spacetimes. At the same opportunity, the possibility of the field-dependent
condensate, such as $\Lambda(\bar{\phi})$, could be found by further study.




\bibliographystyle{apsrev4-1}

\end{document}